\documentclass[aps]{revtex4}
\usepackage{epsfig,amsmath,amssymb,graphics}

\parskip=\medskipamount

\def\be{\begin{equation}}
\def\ee{\end{equation}}
\def\disp{\displaystyle}

\def\H{{\mathbb{H}}}
\def\C{{\mathbb{C}}}
\def\Z{{\mathbb{Z}}}
\def\D{\C^{\star\star}}
\def\slz{PSL(2,{\mathbb{Z}})}

\begin{document}
\title{Conformal Geometry and Invariants of 3--strand Brownian Braids}
\author{Sergei Nechaev$^{1}$ and Rapha\"el Voituriez$^{2}$}
\affiliation{$^{1}$Laboratoire de Physique Th\'eorique et Mod\'eles
Statistiques, Universit\'e Paris Sud, 91405 Orsay Cedex, France}
\affiliation{$^{2}$Laboratoire de Physique Th\'eorique des Liquides,
Universit\'e Paris VI, 4 Place Jussieu, 75252 Paris Cedex, France}
\date{July 2, 2004}

\begin{abstract}
We propose a simple geometrical construction of topological invariants of 3--strand
Brownian braids viewed as world lines of 3 particles performing independent Brownian
motions in the complex plane $z$. Our construction is based on the properties of
conformal maps of doubly-punctured plane $z$ to the universal covering surface. The
special attention is paid to the case of indistinguishable particles. Our method of
conformal maps allows us to investigate the statistical properties of the
topological complexity of a bunch of 3--strand Brownian braids and to compute the
expectation value of the irreducible braid length in the non-Abelian case.
\end{abstract}

\maketitle

\section{Introduction}
\label{sect1}

Our paper presents some results concerning geometry and statistics of random
three-strand braids. Besides the relevance of this work at a purely algebraic level
(see, for example, \cite{birman} for a review of principal topological problems),
its usefulness in the understanding statistical physics of entangled lines,
independent on their physical nature, is also undeniable. Statistics of ensembles of
uncrossible linear objects with topological constraints has very broad application
area ranging from problems of self-diffusion of directed polymer chains in flows and
nematic-like textures to dynamical and topological aspects of vortex glasses in
high--$T_{\rm c}$ superconductors \cite{nel,nelson}. In order to have representative
and physically clear image for the system of fluctuating lines with non-Abelian
(i.e. noncommutative) topology we formulate the model in terms of entangled Brownian
trajectories: such representation serves also as a geometrically clear image of
Wilson loops in (2+1)D non-Abelian field--theoretic path integral formalism. In
particular our paper focuses on the geometrical, algebraic and statistical
properties of the simplest non-commutative braid group $B_3$.

Consider the ensemble of $N$ particles randomly moving in the complex plane
$z=x+iy$. Let us label the coordinates of these particles by $z_j(t)=x_j(t)+iy_j(t)$
where $1\le j\le N$ and $t$ is the current "time" (the initial configuration of the
particles corresponds to $t=0$). It is clear that in (2+1)D "space-time" $(z,t)$ the
diffusive motion of all particles is described by the statistics of $N$ "world
lines" or "directed polymers" and the time $t$ plays the role of the length of the
"world line". In what follows we shall assume the periodic boundary conditions, i.e.
we shall suppose that at some time $t=T$ the configuration of the particles in the
plane $z$ is just the permutation of the initial configuration, i.e. $\{z_1(T),...,
z_N(T)\}={\mathcal P}\{z_1(0),..., z_N(0)\}$, ${\mathcal P}$ being a permutation of
$N$ elements. Thus, under the natural condition $z_j(t)\neq z_k(t)$ for any $1\le
\{k,j\}\le N$, the "world lines" shall be entangled.

The topology of the braid can play a crucial role in macroscopic physical
properties. Let us mention only two examples. The elastic properties of polymer
networks strongly depend on the initial degree of entanglement between subchains
forming the sample and, hence, the elastic properties of the rubbers can be
controlled by different initial conditions of samples preparation \cite{edwards,
florykh}. In ${\rm Cu0_2}$--based high--$T_{\rm c}$ superconductors in fields less
than the critical magnetic field $H_{\rm c2}$ there exists a region where the
Abrikosov flux lattice is molten, but the sample of the superconductor demonstrates
the absence of the conductivity. This effect is explained by highly entangled state
of flux lines due to their topological constraints \cite{nel,obrub}.

\begin{figure}[ht]
\begin{center}
\epsfig{file=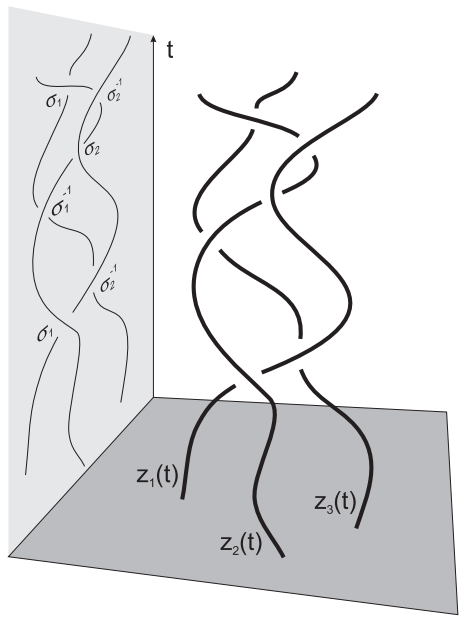,width=6cm}
\end{center}
\caption{Braiding of three directed lines. Elementary moves $\sigma_i$ in the
projection generate the group $B_3$.}
\label{fig:1}
\end{figure}

We restrict our study to $N=3$, i.e. to the case of 3--strand braids produced by
bunches of "word lines" of 3 particles simultaneously moving in the complex plane
$z$ (see fig.\ref{fig:1}). The simplest non-commutative group $B_3$ is defined on
the projection of this bunch onto the plane $(t,x)$ as it is shown in
fig.\ref{fig:1}. The group $B_3$ is constituted by the set of generators
$\{\sigma_1,\sigma_2, \sigma_1^{-1}, \sigma_2^{-1}\}$ satisfying commutation
relations explicited below. Each topological configuration can be exactly encoded by
an element of the braid group $B_3$, that is a "word" written in terms of
"letters"--the braid group generators, associated to elementary
moves--positive/negative crossings in the projection (see fig.\ref{fig:1}). A
configuration of the braid corresponds at the time $t=T$ to an element $\omega_T$ of
$B_3$.

We are interested in an explicit construction of topological invariants of
entanglements of such world lines i.e., in the  $3$--strand  random "braid"
$\omega_T$. We investigate and evaluate the complexity of such randomly generated
braid, defined as the minimal number of generators $L(\omega_T)$ necessary to write
$\omega_T$. This quantity  $L(\omega_T)$ is called the irreducible length  (in the
metric of words) and  exactly coincides with the minimal number of crossings
necessary to represent the entanglement. Therefore $L(\omega_T)$ can be chosen as an
indicator of the braid complexity. In particular, if $L(\omega_t)=0$ the braid is
trivial (i.e. the strands are unentangled).

In what follows we shall repeatedly use the so-called Burau matrix representation of
the braid group. In the particular case of the group $B_3$ the Burau representation
is given by $2\times 2$ matrices:
\be
\sigma_1=\left(\begin{array}{cc} -t & 1 \\ 0 & 1
\end{array}\right); \qquad
\sigma_2=\left(\begin{array}{cc} 1 & 0 \\ t & -t
\end{array}\right) \label{eq:4}
\ee
where $t$ is a free parameter. It is known that for $B_3$ this representation is
faithful \cite{birman}. When $t=-1$ the group $B_3$ coincides with the modular group
$\slz$.

The paper is structured as follows. The first part introduces the basic notions of
homotopy groups and homotopic length, necessary to self-contained description of
topological invariants. The method proposed in this part, demonstrates on the
simplest examples how the monodromy representations of groups allows one to build
the universal coverings, giving rise to topological invariants, associated at a
physical level to fluxes of non-Abelian extension of "solenoidal magnetic fields".
Using the same ideas, the second part deals with the case of $B_3$ for which we
explicitly derive a topological invariant, directly linked to the irreducible braid
length, and construct its non-Abelian flat connection. For the model of ideal
Brownian braids, the stochastic evolution of this topological invariant is given and
its asymptotic distribution is derived.

\section{Topological invariants and monodromy representations}
\label{sect:r-h}

\subsection{Basic concepts and definitions}

Let us consider the double--punctured complex plane $\D={\C}-\{z_1,z_2\}$, of
variable $z=x+i y$ and suppose the coordinates of the punctures $M_1$ and $M_2$ to
be $z_1=(0,0)$ and $z_2=(c,0)$ correspondingly. We set $c\equiv 1$ by appropriate
rescaling of the plane $z$.

Take now two closed elementary paths on $z$ such that the first one ($\gamma_1$)
encloses only the point $M_1$ and the second one ($\gamma_2$) surrounds only the
point $M_2$. With the usual composition law of paths, $\gamma_1$ and $\gamma_2$,
generate a fundamental group $F_2$, which is the first homotopy group, $\pi_1$, of
the double--punctured complex plane $z$:
$$
\pi_1(\D)=F_2
$$
The trivial path (the unit element of the group $F_2$) is the composition of an
arbitrary loop with its inverse: $e=\gamma_i\gamma_i^{-1}=\gamma_i^{-1} \gamma_i,\;
i=\{1,2\}$. The loops $\gamma_i$ and $\tilde{\gamma}_i$ are equivalent if $\gamma_i$
can be continuously deformed into $\tilde{\gamma}_i$ (the equivalent loops represent
the same element of $F_2$). We denote the class of equivalent paths by $\gamma_i$.

Any element of the group $F_2$ being finitely generated, corresponds to a closed
path on $z$ and can be represented by a "word" consisting of a sequence of
letters--generators of the group $F_2$: $\{\gamma_1,\gamma_2,
\gamma_1^{-1},\gamma_2^{-1}\}$. Each word can be reduced to a minimal (or
"irreducible") representation. For example, the word $W=\gamma_1\gamma_2^{-1}
\gamma_1\gamma_1 \gamma_1^{-1}\gamma_2^{-1}\gamma_2\gamma_1^{-1}\gamma_2^{-1} \equiv
\gamma_1\gamma_2^{-1} \left[\gamma_1\left[\gamma_1
\gamma_1^{-1}\right]\left[\gamma_2^{-1}
\gamma_2\right]\gamma_1^{-1}\right]\gamma_2^{-1}$ can be reduced to
$W=\gamma_1\gamma_2^{-1}\gamma_2^{-1}$. Consider now a finitely generated group
$G=\langle g_1,g_2\rangle$ (to be more precise, the group admitting only one-- or
two--generators presentation). We are interested in the monodromy representation of
$F_2$ into $G$ defined by the following group homomorphism $\Psi$:
\be
\Psi:\left\{\begin{array}{lll}
F_2 & \longrightarrow & G \\
\gamma_i & \longrightarrow & g_i \quad (i=1,2)
\end{array}\right.
\ee
Note that this straightforwardly can be generalized to the case of multi--punctured
surfaces or other topological spaces. We do not discuss in details the existence of
such homomorphism, assuming that it results mainly from the fact that $F_2$ is a
free group.

Our main goal in the present paper consists in defining an explicit geometrical
construction of a non-Abelian generalization of a Gauss topological invariant (i.e.
a "linking number") for different groups $G$  using the monodromy representations of
$G$. In particular, we construct a complex flat connection $A(z)$ on $\D$, i.e. a
"Bohm-Aharonov--like vector potential" $A(z)$, whose holonomy gives rise to a
representation of $G$. A special attention is paid to the case when $G=B_3$.
Moreover, we use the developed approach to estimate the averaged complexity of a
3--strand braid represented by independent motion of 3 Brownian particles in (2+1)
dimensions.

\subsection{Topological invariants from conformal maps}

In order to explain the basic notions, we begin with the simplest possible
case---the Abelian group $\Z$ and construct the corresponding topological invariant
by means of conformal transforms.

\subsubsection{Abelian case: commutative group $G=\Z$}
\label{commut}

The central point of the approach deals with the reconstruction of a linear
differential equation on the manifold $\D$ by its monodromy group $G$. In other
words, defining the action of $G$ on the space of solutions of some $2^{\rm
nd}$--order liner differential equation with two branching points, we are attempting
to recover the form of this differential equation and its solutions. In the case of
$G=\Z$ it is natural to consider a usual 1-dimensional additive action
$$
n[w(z)]=w(z)+2i\pi n
$$
for $w$ defined on $\D$ and $n\in\Z$. It turns out that the simplest linear
differential equation with two branching points, satisfying those conditions is:
\begin{equation}
(z-z_1)(z-z_2)\frac{dw}{dz}-2z+z_1+z_2=0
\end{equation}
whose solution reads
\begin{equation}
w(z)=\ln(z-z_1)+\ln(z-z_2) \label{eq:commut}
\end{equation}
One can check that the free group $\Gamma_2$ acts on the space of solutions as follows
\begin{equation}
w(z)\stackrel{\{\gamma_1,\gamma_2\}}{\longrightarrow} w(z)+2i\pi \label{action}
\end{equation}
which means that the homomorphism $\Psi$ is trivial in for this example:
$\Psi(\gamma_i)=1$ $(i=1,2)$. The function $w(z)$ conformally maps the doubly
punctured plane to the universal covering space $w=u+iv$ free of any branching
points. In the complex plane $w$ we have
$$
\left\{\begin{array}{l}
u=\ln|(z-z_1)(z-z_2)| \medskip \\ v=\arg (z-z_1)+ \arg (z-z_2)
\end{array}\right.
$$
The function $w(z|z_1,z_2)$ is a topological invariant because of equality
(\ref{action}). In particular, the function $v(z|z_1,z_2)$ defines the total number
of turns in the plane $z$ around the branching points $z_1$ and $z_2$ and hence it
is nothing else as the Gauss linking number.

Thus, knowing the conformal transform $w(z)$ of the multiply punctured plane $\D$ to
the universal (i.e. uniformizing) covering surface, we can easily extract a
topological invariant ${\rm Inv}(\gamma)$ of a closed path $\gamma$ (starting and
ending at some arbitrary point $z_0\neq \{0,1\}$ in the plane $z$) from the
difference $w_{\rm in}(z_0)-w_{\rm fin}(z_0)$. The path $\gamma$ connects the images
of the point $z_0$ on different Riemann sheets---the "copies" of the fundamental
domain of $G$. Recall that by definition the fundamental domain of a group $G$ is a
minimal connected domain tessellating the whole covering space under the action of
$G$. Representing the topological invariant ${\rm Inv}_{(z)}(\gamma)$ as a full
derivative along the contour $\gamma$, we get:
\be
{\rm Inv}_{(z)}(\gamma)\stackrel{\rm def}{=}w_{\rm fin}-w_{\rm in}\equiv
\oint\limits_{\gamma}\frac{dw(z)}{dz} dz = \oint A(z) dz \label{eq:def_top}
\ee
The physical interpretation of the derivative $A(z)=\frac{dw(z)}{dz}$ is very
straightforward. The conformal transform $w(z)$ plays the role of a complex
potential of a field $A(z)$, which defines a {\it flat  connection} of a
multiple--punctured plane ${\C}^{\star\star}$.

For the commutative group $G=\Z$ we obtain $A(z)$ by taking the derivative of
(\ref{eq:commut}):
\be
A(z)=\frac{1}{z-z_1}+\frac{1}{z-z_2} \label{eq:commut2}
\ee
This expression can be easily identified with the standard (Abelian) Bohm--Aharonov
vector potential of two solenoidal magnetic fields orthogonal to the plane $z$ and
crossing it in the points $z_1$ and $z_2$.

In the next section the same construction of the flat connection associated to a
specific group $G$ will be generalized to the non-Abelian case.

\subsubsection{Non-Abelian case: non-commutative groups $F_2,\, H_q$}
\label{noncommut}

We consider a special class of hyperbolic and  hyperbolic--like groups: the free
($F_2$) and the Hecke ($H_q$) groups, as well as the braid group $B_3$. By
hyperbolic--like groups we mean a class broader than hyperbolic groups in the
classification of M.Gromov \cite{gromov}.  (According to the Gromov's definition,
the group $B_3$ does not belong to the class of hyperbolic groups). The important
feature for us would be just the exponential growth of the group. From this point of
view the group $B_3$ fits our scheme.

\noindent 1. The free group $F_2$ (in general, $F_n\; (n=2,3,4,...)$) by definition
is the free product of two (in general of $n$) copies of cyclic groups of second
order, ${\Z}_2$. The matrix representation of the generators of the free groups
$F_2\{f_1,f_2\}$ and $F_3\{f_1,S\}$ are:
\be
\begin{array}{l}
F_2: \quad f_1=\left(\begin{array}{cc} 1 & 2 \\ 0 & 1
\end{array}\right);\;
f_2=\left(\begin{array}{cc} 1 & 0 \\ 2 & 1
\end{array}\right) \medskip \\
F_3: \quad f_1=\left(\begin{array}{cc} 1 & 2 \\ 0 & 1
\end{array}\right); \;
S=\left(\begin{array}{cc} 0 & 1 \\ -1 & 0
\end{array}\right)
\end{array}
\ee

\noindent 2. The Hecke group $H_q$ is the free product of two cyclic groups $\Z_2$,
and $\Z_q$ of orders 2 and $q$ respectively. The Hecke group is defined by the
relations
\be
\begin{array}{r}
(ST_q)^q=b_{q}^q=1 \medskip \\ S^2=a_{2}^2=1
\end{array}
\ee
where the generators $T_q$ and $S$ have the following matrix representation
\be
H_q: \quad T_q=\left(\begin{array}{cc} 1 & 2\cos\frac{\pi}{q} \\ 0 & 1
\end{array}\right);\; S=\left(\begin{array}{cc} 0 & 1 \\ -1 & 0
\end{array}\right)
\ee
The parameter $q$ takes discrete values $q=3,4,5,6,\ldots$. The Hecke group $H_q$
"interpolates" between the modular group $\slz$ (for $q=3$) and the free group $F_3$
with 3 generators ($q\to \infty$).

We have stressed in the previous section that the topological invariant can be
constructed on the basis of the conformal map $w(z)$ of multiple--punctured plane to
the universal covering space of a group. We now extend the described method to more
interesting cases than considered at length of the section \ref{commut}.

\begin{figure}[ht]
\begin{center}
\epsfig{file=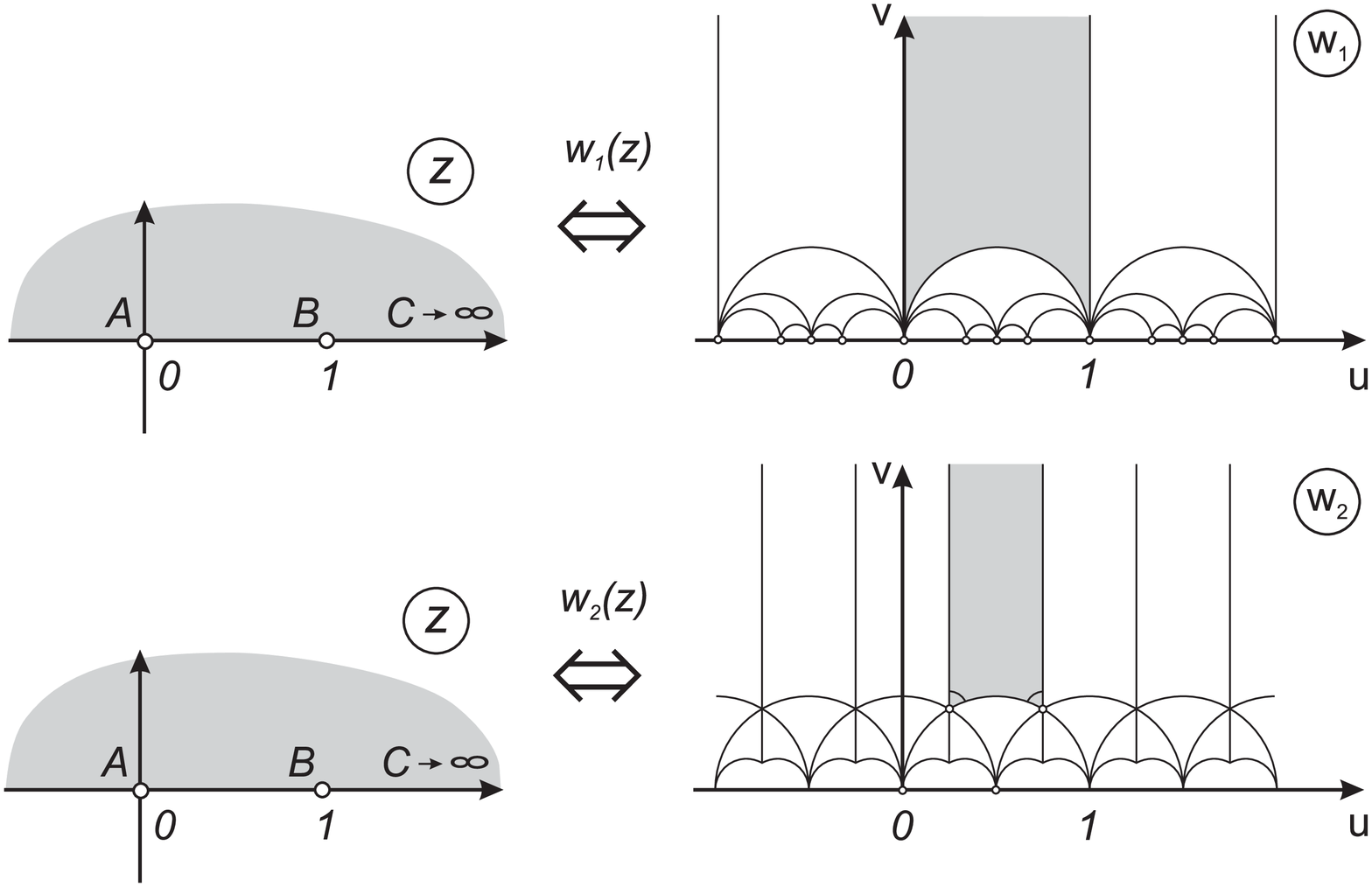,width=12cm}
\end{center}
\caption{Conformal transforms of the upper half-plane ${\rm Im}\,z>0$ to the
fundamental domain of the group $F_2$ (up) and of the group ${\slz}$ (down).}
\label{fig:2}
\end{figure}

We derive the conformal mapping of the half--plane ${\rm Im}\,z>0$ onto the
fundamental domain of the triangular group $G$---a curvilinear triangle lying in the
upper half--plane ${\rm Im}\,w>0$. The action of $G$ on this fundamental domain
generates the whole covering space. Each copy of the fundamental domain represents a
Riemann sheet corresponding to the fibre bundle above $z$ and the whole covering
space $w$ is the unification of all such Riemann sheets---see fig.\ref{fig:2}.

The coordinates of initial and final points of a trajectory on the universal
covering $w$ determine:
\begin{itemize}
\item The coordinates of the corresponding points on $z$;
\item The element of $G$ corresponding to the homotopy class of the path on $z$.
\end{itemize}

Following the construction described in the previous section let us define the
action of $G\{g_1,g_2\}$ in the complex plane $w=u+iv$. The group $G$ admits the
faithful 2--dimensional representations and acts in the covering space $w$ by
fractional--linear transforms. Consider two basic contours $\gamma_1$ and
$\gamma_2$, associated to the action of $G$ on $z$ ($z\neq \{z_1,z_2,\infty\}$. The
contours $\gamma_1$ and $\gamma_2$ enclose the branching points located at $z_1$ and
$z_2$ correspondingly ($z\neq \{z_1,z_2,\infty\}$). The function $w(z)$ ($z\neq
\{z_1,z_2,\infty\}$) obeys the following transformations:
\be
w\left[z\stackrel{\gamma_1}{\rightarrow} z\right]\rightarrow
\tilde{w}_1(z)= \frac{a_1 w(z)+b_1}{c_1 w(z)+d_1}; \quad
w\left[z\stackrel{\gamma_2}{\rightarrow} z\right]\rightarrow
\tilde{w}_2(z)= \frac{a_2 w(z)+b_2}{c_2 w(z)+d_2} \label{monodr}
\ee
The matrices $g_1$ and $g_2$
\be
g_1=\left(\begin{array}{cc}a_1 & b_1 \\ c_1 & d_1 \end{array}\right);
\qquad g_2=\left(\begin{array}{cc}a_2 & b_2 \\ c_2 & d_2 \end{array}\right)
\ee
are the matrices of basic substitutions of the group $G\{g_1,g_2\}$, i.e. $g_1$ and
$g_2$ are the generators of $G$.

It is well known that the function $w(z)$ can be defined as the quotient of two
fundamental solutions $u_1(z)$ and $u_2(z)$ of a second order differential equation
with branching points $\{z_1=(0,0), z_2=(0,1), z_3=\infty\}$. As it follows from the
analytic theory of differential equations, the solutions $u_1(z)$ and $u_2(z)$
undergo the linear substitutions when the variable $z$ moves along the contours
$\gamma_1$ and $\gamma_2$:
\be
\gamma_1:\,\left(\begin{array}{c}\tilde{u}_1(z) \\ \tilde{u}_2(z)
\end{array}\right) = g_1 \left(\begin{array}{c} u_1(z) \\
u_2(z) \end{array}\right);\quad
\gamma_2:\,\left(\begin{array}{c}\tilde{u}_1(z) \\ \tilde{u}_2(z)
\end{array}\right) = g_2 \left(\begin{array}{c} u_1(z) \\
u_2(z) \end{array}\right) \label{monomat}
\ee
where $g_1$ and $g_2$ are the generators of the monodromy group of this equation.
The problem of restoring the form of a differential equation by the monodromy
matrices $g_1$ and $g_2$ of the group $G$ of the differential equation, is known as
the Riemann-Hilbert problem. Yet we restrict ourselves with the groups $F_2$ and
$H_q$, the group $B_3$ shall be considered separately.

\bigskip

\noindent{\sc The free group $F_2$.} For the free group $F_2$ the solution of the
Riemann-Hilbert problem gives rise to the following second--order differential
equation \cite{bateman}:
\be
z(z-1)\frac{d^2u(z)}{dz^2}+(2z-1)\frac{du(z)}{dz}+\frac{1}{4}u(z)=0 \label{eq:free}
\ee
Indeed, a possible basis of solutions of this equation is as follows:
\be
\left\{\begin{array}{ll}
u_1(z)=F(1/2,1/2,1,z) \medskip \\
u_2(z)=iF(1/2,1/2,1,1-z)
\end{array}
\right. \label{eq:free-sol}
\ee
Using the well--known properties of hypergeometric functions, one can restore the
monodromy matrices defined in (\ref{monomat}) for this basis:
\be
g_1=\left(\begin{array}{cc}1 & 0 \\ 2 & 1 \end{array}\right);
\qquad g_2=\left(\begin{array}{cc} 1 & -2 \\ 0 & 1 \end{array}\right)
\ee
which coincides with the generating set of the group $F_2$. The function $w(z)$
performing the conformal map of the upper half--plane Im$z>0$ onto the fundamental
domain (the curvilinear triangle $ABC$) of the universal covering $w$ satisfies
eq.(\ref{monodr}) and can be written as:
\be
w(z)=\frac{u_1(z)}{u_2(z)} \label{map}
\ee
The function $w(z)$ is well known in the literature (see, for example,
\cite{bateman})  and its inverse $z(w)$ is the elliptic modular function
\be
z(w)\equiv \lambda(w)=\frac{\theta_2^4(0,w)}{\theta_3^4(0,w)} \label{eq:modular}
\ee

Now we can give an explicit expression of the flat connection ${\bf A}(z)$ for the
doubly punctured plane corresponding to the monodromy of the free group  $F_2$.
Taking the derivatives and using the properties of the hypergeometric functions, we
get:
\be
A(z)=\frac{dw(z)}{dz}=-i\frac{E(z)K(1-z)+(E(1-z)-K(1-z))K(z)}{2z(z-1)K^2(1-z)}
\label{eq:conn}
\ee
where
\be
K(z)=\int\limits_0^{\pi/2}\frac{d\theta}{\sqrt{1-z\sin^2\theta}};\qquad
E(z)=\int\limits_0^{\pi/2} d \theta \sqrt{1-z\sin^2\theta}
\label{eq:conn1}
\ee
are correspondingly the complete elliptic integrals $K(z)$ and $E(z)$ (see, for
example, \cite{abramowitz}).

\begin{figure}[ht]
\epsfig{file=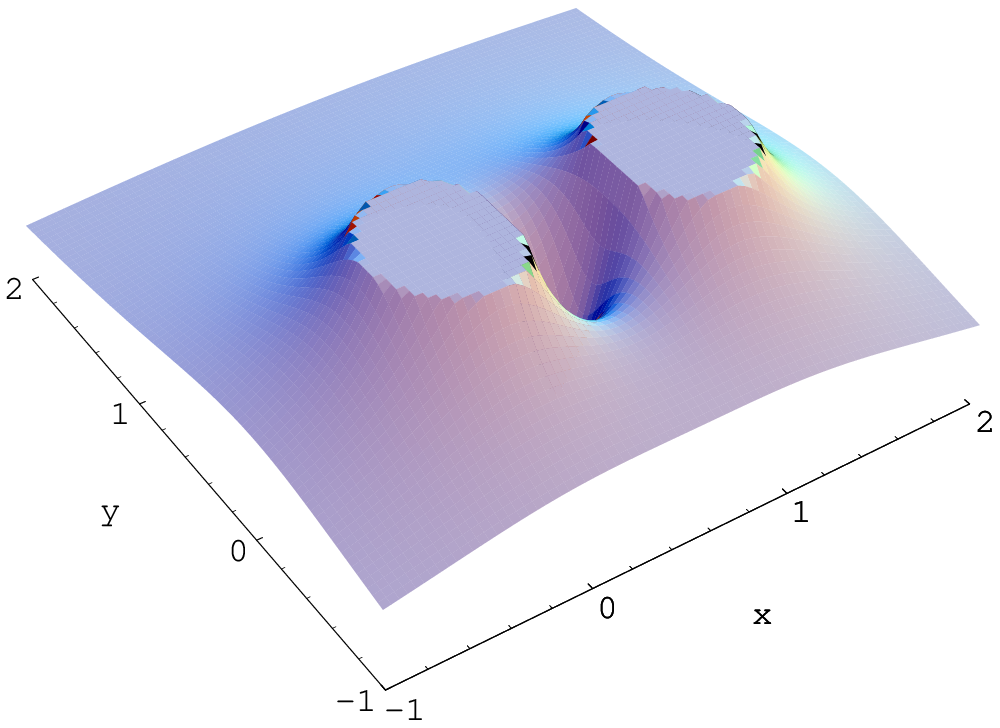,width=7cm} \hspace{1cm} \epsfig{file=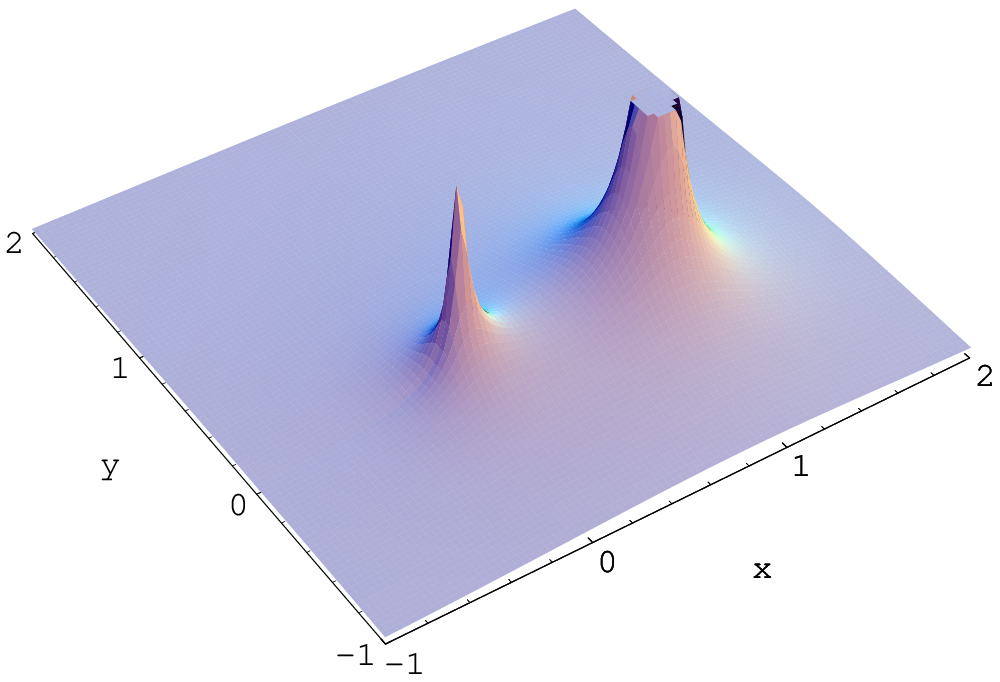,width=7cm}
\caption{The cuts of the function $|A(z)|$ in the Abelian (left) and non-Abelian
(right) cases.} \label{fig:3}
\end{figure}

In the Fig.\ref{fig:3} we have plotted the absolute values $|A(z)|$ of the Abelian
(Eq.(\ref{eq:commut2})) and the non-Abelian (Eq.(\ref{eq:conn})) flat connections.

The leading asymptotics of (\ref{eq:conn}) are as follows:
\be
A(z)\sim \left\{\begin{array}{cl}
\disp \frac{-i\pi}{z\ln^2 z} & \quad \mbox{as $z\to 0$} \medskip \\
\disp \frac{-i}{\pi(z-1)} & \quad \mbox{as $z\to 1$}
\end{array}\right. \label{eq:asympt}
\ee
Hence even in the vicinity of the branching point $z=0$ the function $A(z)$ defined
in (\ref{eq:conn}) does not coincide asymptotically with the flat connection for the
Abelian case (\ref{eq:commut2}).

\bigskip

\noindent{\sc The Hecke group $H_q$.} We now pass to the case of $H_q$, keeping in
mind that for $q=3$ we recover the group $\slz$, directly linked to $B_3$ ($B_3$ is
a central extension of $\slz$, see \cite{birman}).  The solution of the
corresponding Riemann-Hilbert problem leads to the following differential equation:
\cite{venkov}
\be
z(z-1)\frac{d^2u(z,q)}{dz^2}+\left[\left(\frac{3}{2}-\frac{1}{q}\right)z-
\left(1-\frac{1}{q}\right)\right]\frac{du(z,q)}{dz}
+\frac{1}{4}\left(\frac{1}{2}-\frac{1}{q}\right)^2 u(z,q)=0 \label{eq:slz1}
\ee
whose possible basis of solution is:
\be
\left\{\begin{array}{ll} \disp
u_1(z)=F\left(\frac{q-2}{4q},\frac{q-2}{4q},1-\frac{1}{q},z\right)-
\lambda(q)e^{i\pi/q}z^{1/q}F\left(\frac{q+2}{4q},\frac{q+2}{4q},1+\frac{1}{q},z\right)
\medskip \\ \disp
u_2(z)=-e^{i\pi/q}F\left(\frac{q-2}{4q},\frac{q-2}{4q},1-\frac{1}{q},z\right)+
\lambda(q)z^{1/q}F\left(\frac{q+2}{4q},\frac{q+2}{4q},1+\frac{1}{q},z\right)
\end{array} \right. \label{eq:slz2}
\ee
where
\be
\lambda(q)=\disp e^{2i\pi/q}\frac{\Gamma(1-\frac{1}{q})\Gamma(\frac{q+2}{4q})
\Gamma(\frac{3q+2}{4q})}{\Gamma(1+\frac{1}{q})\Gamma(\frac{q-2}{4q})
\Gamma(\frac{3q-2}{4q})} \label{eq:slz3}
\ee
Taking into account that the series defining the hypergeometric functions $F(z)$
converges for $|z|<1$ and corresponds to the so-called logarithmic case (see
\cite{bateman}) one obtains from (\ref{eq:slz2}) the following monodromy matrices
after proper analytic continuation:
\be
g_1=e^{i\pi/q}\left(\begin{array}{cc} 2\cos\frac{\pi}{q} & 1 \\ -1 & 0
\end{array}\right);
\qquad g_2=i\left(\begin{array}{cc} 0 & -1 \\ 1 & 0 \end{array}\right) \label{eq:slz4}
\ee
One can check that after normalization by the determinant, the matrices
(\ref{eq:slz4}) become the generators of $H_q$ (correspondingly of orders $q$ and
2). Let us point out that for $q=3$, one can identify the inverse function $z(w)$
with the Klein's "absolute modular invariant"
$$
z(w)\equiv J(w)=\frac{(\theta_2^8(0,w)+\theta_3^8(0,w)+\theta_4^8(0,w))^3}
{54\theta_2^8(0,w)\theta_3^8(0,w)\theta_4^8(0,w)}
$$
where $\theta_2,\theta_3,\theta_4$ are the  Jacobi elliptic $\theta$--functions. As
in the previous section, the corresponding complex flat connection $A_{H_q}(z)$ can
be obtained as the full derivative
$$
A_{H_q}(z)=\frac{dw_{H_q}(z)}{dz}
$$
from the conformal map
$$
w_{H_q}(z)=\frac{u_1(z)}{u_2(z)}
$$
with $u_1(z)$ and $u_2(z)$ defined in (\ref{eq:slz2}).

\section{Invariant of 3-strand braids}
\label{sect:3}

Let us return to the model of random braids discussed in the very beginning of the
paper.  The system of three independently moving particles is described by 3 complex
variables (i.e. 6 degrees of freedom) and hence our system can not be directly
viewed as a monodromy problem. In order to describe the whole system by one complex
variable $\zeta$ and to be able to use the tools elaborated in the previous section,
with the minimal loss of information, we introduce the anharmonic quotient:
\be
\zeta(t)=\frac{z_{23}(t)}{z_{13}(t)}\equiv \frac{z_2(t)-z_3(t)}{z_1(t)-z_3(t)}
\label{eq:anh}
\ee
One can easily check that the variable $\zeta$ contains the complete topological
information of a mutual configuration of three entangled lines, except the global
phase, or global twist of the braid. It means that we do not take into account the
center of $B_3$ considering the factor group ${\slz}=B_3/{\Z}$. We can always take
into account the global twist afterwards by passing to the rotating coordinate
system with the origin located in the center of mass of the system of three
particles. From the statistical point of view the possibility of neglecting the
center of $B_3$ (which is equivalent to considering the factor group $\slz$ only),
has been discussed in the papers \cite{raf,voitnech3}. In fact, in these paper it
has been shown that the escape rates for random walks on $B_3$ and $\slz$ are the
same in the limit of infinitely long trajectories.

Thus, the parametrization of the three-particle system
$\{z_1(t),\,z_2(t),\,z_3(t)\}$ by the function $\zeta(t)$ living in the doubly
punctured complex plane enables us to preserve all topological characteristics of
the braid of mutually entangled world lines of these particles. However the precise
derivation of the expression for the flat connection, as it is shown below,
explicitly depends on the fact whether the particles are identical or not.

Looking at the elementary moves associated with the generators $\sigma_1,\ \sigma_2$
of $B_3$, we obtain the transformations for the variable $\zeta$. For example, let
us pick up some path in the homotopy class of $\sigma_1$ (see fig.\ref{fig:1}). By
definition of $\sigma_1$, we have at some time $t$:
\be
\sigma_1[z_1\leftrightarrow z_2, z_3\leftrightarrow z_3]:
\left\{\disp \begin{array}{lll}
\disp z_1(t)=\frac{z_1+z_2}{2}+\frac{1}{2}z_{12}e^{-i\pi t} \medskip \\
\disp z_2(t)=\frac{z_1+z_2}{2}-\frac{1}{2}z_{12}e^{-i\pi t} \medskip \\
z_3(t)=z_3 \end{array}\right. \label{eq:b3}
\ee
In the same way we can get the transformation of $\zeta$ along a path corresponding
to the homotopy class $\sigma_2\equiv \sigma_2[z_1\leftrightarrow z_1, z_2
\leftrightarrow z_3]$. Finally we arrive at the following set of transformations of
$\zeta$:
\be
\begin{array}{lcc}
\disp \sigma_1:\,  \zeta & \to & \disp \frac{1}{\zeta} \medskip \\
\disp \sigma_2:\,  \zeta & \to & \disp \frac{\zeta}{\zeta-1}
\end{array} \label{eq:symmet}
\ee
which precisely define the representation of the symmetric group $S_3$.

\bigskip

\noindent {\sc Indistinguishable particles.} Our main goal consists in constructing
the flat connection for the system of three {\it identical} particles moving in the
plane. The condition of the indistinguishability of particles requires to factor the
action of the group $B_3$ by $S_3$, and hence to consider $\zeta$ as living in the
factor space $\D/S_3$. The case of {\it distinguishable} particles shall be
discussed at the end of this section.

All the trajectories $\zeta(t)$ (parameterized by $t$) are obviously closed in the
space $\D/S_3$. Keeping in mind our strategy of the previous section, we want to
define a monodromy representation of $B_3$ (more precisely of $\slz$) acting on
$\D/S_3$. To do that we construct a conformal map $w_s(\zeta)$ of the doubly
punctured plane factorized over the action of the symmetric group, $\D/S_3$ onto the
fundamental domain of the modular group, $\slz$:
\be
w_s(\zeta):\, \D/S_3 \longrightarrow \slz \label{eq:unif}
\ee

To our knowledge, the explicit method of solving the uniformization problem
(\ref{eq:unif}) has not been yet considered in the context of our problem. So, our
own way to tackle it deals with the following observation.

\begin{figure}[ht]
\begin{center}
\epsfig{file=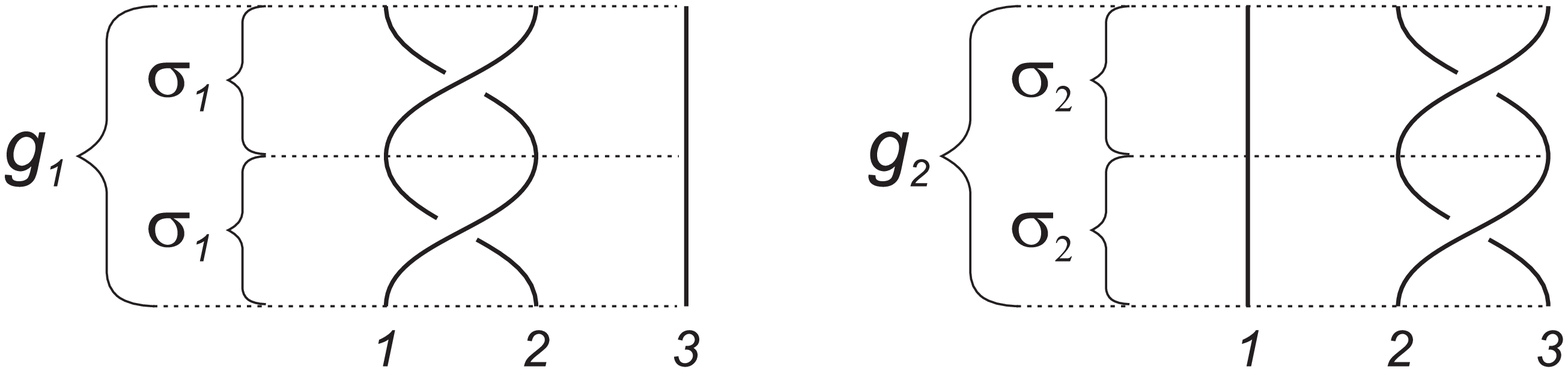,width=10cm}
\end{center}
\caption{The squares of the braid group generators $\sigma_i$ act as the free group
generators $g_i$: $g_i=\sigma_i^2$ ($i=1,2$).} \label{fig:4}
\end{figure}

Consider the so-called pure braid group $P_3$ defined as follows:
\be
P_3=\langle\sigma_1^2,\sigma_2^2,(\sigma_1\sigma_2)^2\rangle
\ee
It is obvious that the group $P_3$ is the subgroup of $B_3$. In particular, the
group $P_3$ can be identified with the factor group $F_2\times\Z$ through the
obvious correspondence
\be
\sigma_1^2=g_1,\; \sigma_1^2=g_1 \label{eq:free1}
\ee
where $g_1$ and $g_2$ are the generators of the free group $F_2$. The relations
(\ref{eq:free}) follow from the Burau representation of braid group generators
(\ref{eq:4}) and the obvious geometric construction shown in the fig.\ref{fig:4}.

In the case of the pure braid group $P_3$ there is no difference between
distinguishable and indistinguishable particles moving in the plane $z$ because the
generators $g_1$ and $g_2$ correspond to "full turns" of one world line of the
particle with respect to another one (and not to "half turns" as in case of the
braid group $B_3$). That is, we deal with the monodromy representation of $F_2$. Let
us remind that we factor out the global turns (i.e. the center of the group) in the
space $\D$. The uniformization problem in this case is solved by the conformal map
$w_s(\zeta)$ of the doubly punctured plane to the covering space
$$
w_s(\zeta): \, \D \longrightarrow F_2
$$
Thus, we return to the topic of the previous section, where an explicit form of such
conformal map $w_s(\zeta)\equiv w(\zeta)$ is given in (\ref{map}).

Consider now the whole group $B_3/\Z$, take the same function $w(\zeta)$ and check
the effect of the basic substitutions of the symmetric group (\ref{eq:symmet}) onto
the transformations of $w(\zeta)$. Using the well known properties of the modular
functions (see, for example, \cite{hille,golubev,mckay}), we get:
\be
\left\{\begin{array}{lcc}
\disp w_s\left(\zeta\to \frac{1}{\zeta}\right) & \to & \disp -w_s+1 \medskip \\
\disp w_s\left(\zeta\to \frac{\zeta}{\zeta-1}\right) & \to & \disp \frac{w_s}{w_s-1}
\end{array} \right. \label{eq:b3-s3}
\ee
The transformations of $w_s$ precisely coincide with the basic transformations of
the modular group $\slz$. In fact, technically it is much easier to check (instead
of (\ref{eq:b3-s3})) the transformations for the inverse function $\zeta(w_s)=
\frac{\theta_2^4(0,w_s)}{\theta{(0,w_s)}}$:
\be
\left\{\begin{array}{lcc}
\disp \zeta\left(w_s\to -w_s+1\right) & \to & \disp \frac{1}{\zeta} \medskip \\
\disp \zeta\left(w_s\to\frac{w_s}{w_s-1}\right) & \to & \disp \frac{\zeta}{\zeta-1}
\end{array} \right. \label{eq:b3-s3-1}
\ee

Hence the function $w_s(\zeta)$ explicitly solves the desired uniformization problem
(\ref{eq:unif}). Thus, we can use this function $w_s(\zeta)$ to construct the
non-Abelian flat connection $A_s(\zeta)$ for the trajectories on $\D/S_3$
parametrized by $\zeta(t)$:
\be
A_s(\zeta)=\frac{dw_s(\zeta)}{d\zeta} \label{eq:as}
\ee
where
$$
w_s(\zeta)\equiv w(\zeta)=\frac{u_1(\zeta)}{u_2(\zeta)}=
-i\,\frac{F(1/2,1/2,1,z)}{F(1/2,1/2,1,1-z)}
$$
(see eq.(\ref{eq:free-sol})).

The expression (\ref{eq:as}) is the source for the explicit construction of the
topological invariant via equation (\ref{eq:def_top}) of three entangled world lines
$z_1(t)$, $z_2(t)$ and $z_3(t)$ shown in fig.\ref{fig:1}.

\bigskip

\noindent {\sc Distinguishable particles.} For {\it distinguishable} particles the
action of the symmetric group $S_3$ should be neglected and hence $\zeta$ is living
in the doubly punctured plane, $\D$. The problem of uniformization in this case is
solved by the conformal map $w_d(\zeta)$ mapping the doubly punctured plane, $\D$,
onto the fundamental domain of the modular group, $\slz$:
\be
w_d(\zeta):\, \D \longrightarrow \slz \label{eq:unif1}
\ee
(compare this expression to (\ref{eq:unif})). The corresponding function
$w_d(\zeta)$ coincides with $w_2(\zeta)=\frac{u_1(\zeta)}{u_2(\zeta)}$ found in the
privious section (use the equations (\ref{eq:slz2})--(\ref{eq:slz3}) for $q=3$).

\section{Stochastic behavior of the invariant for Brownian 3-strand braids}

We are interested in this section in the {\it stochastic} behavior of the invariant
$|w(\eta)|$ when each of the three particles perform a 2D Brownian motion (BM)
represented by its complex coordinate $z_i$. This model naturally appears as the
generalization of the Edwards' problem of determining the distribution of the
winding angle of two independent 2D BM \cite{edwards}. The intrinsic noncommutative
structure of the 3--particles model, contrary to the commutative 2--particles
problem, induces a completely different behavior of the invariant, which
characterizes quantitatively the tangle complexity. The method employed here is
inspired by a work of Gruet \cite{gruet}.

\subsection{The Ideal Model}

We begin with an ideal case of point--like particles diffusing in the infinite plane
with the diffusion constants $D$ set to 1. Following previous sections, one has to
study the motion of the reduced variable $\eta$ in the space $\C^{\star\star}/S_3 $.
Our strategy relies on the following theorem on conformal transforms of complex BM
often referred to as a generalized Paul L\'evy's theorem \cite{yor}. Denote by $B_t$
the generic BM in the complex plane $z$. For any conformal map $f$, the image
$Y_t=f(B_t)$ of a BM $B_t$ is a \textit{time changed} BM $B_u(t)$ such that
\be
du=\dfrac{1}{4}d Y d\bar{Y}=\vert f'(B_t)\vert^{2}dt \label{stoch1}
\ee
In terms of the diffusion equation (which corresponds to the Langevin equation
(\ref{stoch1}), the time change can be interpreted as a space dependent diffusion
coefficient \cite{ito-mckean}:
\be
\partial_t P(Y_t)=\dfrac{du}{dt}\Delta P(Y_t) \label{stoch2}
\ee
A slightly modified version of the generalized L\'evy theorem shows that $\zeta(t)$,
being the ratio of two complex BM (see (\ref{eq:anh})), is also a time changed BM,
denoted $\zeta(t)=B_{u(t)}$. The time change then reads, following the rules of
It\^o calculus:
\be
du=\frac{1}{4}d\zeta d\bar{\zeta}=\frac{1+|\zeta(t)|^2}{|z_{32}(t)|^2}dt
\label{tchange1}
\ee

Now we should pass to the covering space $w$ and describe the evolution of the
invariant $w(\zeta)=w(B_u)$. Using once more the generalized L\'evy's theorem, we
obtain that there exists a {\it hyperbolic} BM $\tilde{B}$ such that $w(\zeta)$ is
again a time--changed $\tilde{B}$:
\be
w(\zeta(t))=\tilde{B}_{\tau(u(t))}
\ee
with
\be
d\tau=\dfrac{|w'(B_u)|^2}{{\rm Im}^2(w(B_u))}du \label{tchange2}
\ee

The choice of the hyperbolic metric is not artificial and anticipates the
geometrical properties of the transform $w(\zeta)$. Note the importance of obtained
result: we know that the topological invariant $w(\zeta)$, giving the braid
complexity, performs a hyperbolic diffusion in a new time, or equivalently a
hyperbolic diffusion in a space with the metric--dependent diffusion coefficient. We
now have to study the full time change $\tau(t)$. Combining (\ref{tchange1}) and
(\ref{tchange2}) and using the well known inverse function $w^{-1}\equiv
z(w)=\lambda(w)$ (see (\ref{eq:modular})), we obtain after separating the variables
the following implicit relation between $t$ and $\tau$, which defines the functional
$H_t(\tau)$:
\be
H_t(\tau)=\int_0^{t}\frac{dt'}{|z_{32}(t')|^{2}}=\int_0^\tau
\frac{|\lambda'(\tilde{B}_{\tau'})|^2}{1+|\lambda(\tilde{B}_{\tau'})|^2}{\rm
Im}^2(\tilde{B}_{\tau'})d\tau' \label{integ}
\ee
This expression is exact, but does not allow a straightforward interpretation. Let
us first notice that the $t$--dependence can be easily shown to be, with the
probability one:
\be
\lim_{t\to \infty}\frac{H_t(\tau)}{\ln t}=C_1>0 \label{lim1}
\ee
The sensitive point is the $\tau$--dependence. The integral kernel $f$
\be
f:\;z\longrightarrow \frac{|\lambda'(z)|^2}{1+|\lambda(z)|^2}{\rm Im}^2(z)
\ee
is invariant under the transformations of the group $F_2$, and $H_t(\tau)$ is
therefore an additive functional of the BM on the quotient surface $\H/F_2$. This
motion is ergodic. Therefore if $H_t(\tau)$ is integrable over the fundamental
domain of $F_2$, one can clear up the asymptotic $\tau$--dependence (see in
particular a similar approach proposed by Gruet in \cite{gruet}). In our case
$H_t(\tau)$ is not integrable over the domain of $F_2$; the $\tau$--dependence for
this ideal model requires more attention and have to be treated with more care. We
expect to consider that question in details separately, while below we express a
conjecture based on the fact that the non-integrability does not lead to a growth
faster than $\tau$. We indeed believe that the non-integrability gives rise to
sub-leading terms in $\tau$ and the following limit holds:
\be
\lim_{\tau\to \infty}\frac{H_t(\tau)}{\tau}=C_2 \label{lim2}
\ee
Comparing (\ref{lim1}) and (\ref{lim2}) we arrive at the relation
\be
\lim_{t\to\infty} \frac{\tau(t)}{t}=\frac{C_1}{C_2} \label{lim12}
\ee

Using (\ref{lim12}) we formulate a central result connecting the hyperbolic BM
$\tilde{B}_\tau$ describing the connection of the expectation of our topological
invariant $d(w)$ being the distance in the covering space $w$, with the word length
$L(\tilde{B}_\tau)$ of the corresponding random braid. Recall that the distribution
of $\tilde{B}_\tau$ is well known, and in particular,
$$
\lim_{\tau\to\infty}\frac{d(\tilde{B}_\tau)}{\tau}=1
$$
almost surely. Moreover, if one notices that the word lengths $L$ for $\slz$ and
for $F_2$ are quasi-isometric, we have:
\be
\lim_{\tau\to\infty}\frac{\left<L(\tilde{B}_\tau)\right>}{\tau\ln \tau} = C_3>0
\quad \mbox{in probability}
\ee
what is a direct consequence of a theorem by Gruet proved in \cite{gruet}. Combining
this result with (\ref{lim12}, we end up with the following leading asymptotics at
large time for the expectation of the word length of Brownian 3--strand braids:
\be
\left\langle L_t\right\rangle \sim \ln(t)\times\ln(\ln t) \quad \mbox{for
$t\to\infty$}
\ee
The average topological complexity increases with time, however the growth is very
slow. It is limited (as in the case of two particles where this logarithmic scaling
also holds) by the angular part of the 2D BM in an infinite space
\cite{ito-mckean,edwards2}: when two particles are far apart (and the corresponding
typical distance grows with time), their relative angle varies very slowly.

\subsection{Model with a compact domain}

We here discuss the physical reasons that justify a regularization of the integral
(\ref{integ}) over the fundamental domain of $F_2$. The divergence of this integral
occurs when $\eta$ approaches the points $\{0,1,\infty\}$ meaning that either two
particles collide, or one particle goes to infinity. It is then quite natural to
modify slightly the model in order to avoid these pathological cases, which in fact
are not realistic if one describes the physical objects of finite thickness such as
polymers. Specifically, we assume that three Brownian particles evolve in a bounded
domain of characteristic size $R$ and that they experience hard-core repulsion at
distance $r$. This can be re-expressed in terms of constraints due to hard walls
added to the domain of $\eta$, hereafter denoted by $D$:
\be
\begin{array}{ll}
\epsilon(r,R)\le |\eta| \le M(r,R) \medskip \\
|\eta(r,R)-1|\ge \epsilon
\end{array}
\ee

The precise shape of the boundary of $D$ is not important. Note that $w(D)$ is
compact in $\H$. Restricting our model to this domain, we can now claim the
integrability of the functional $H^{D}_t(\tau)$ over the domain $w(D)$. Within this
approximation, we obtain the following $\tau$--dependence from the ergodic property
offered above:
\be
\lim_{\tau\to \infty}\frac{H^{D}_t(\tau)}{\tau}=C_4>0 \label{lim4}
\ee
In this model the $t$--dependence is readily changed as follows:
\be
\lim_{t\to \infty}\frac{H^{D}_t(\tau)}{t}=C_5>0 \label{lim5}
\ee
Comparing (\ref{lim4}) and (\ref{lim5}) we arrive at
\be
\lim_{t\to\infty}\frac{\tau(t)}{t}=\frac{C_4}{C_5} \label{lim45}
\ee

The expectation of $d(w)$ can now be described asymptotically: $d(w)$ is a
hyperbolic BM in the variable $\tau\propto t$ (see (\ref{lim45})). In particular, we
have with the probability one:
\be
\lim_{t\to \infty}\frac{\left<d(w(\eta))\right>}{t}=C_6>0
\ee
For a compact domain one can straightforwardly check that the topological invariant,
$d$, and the word length, $L$, are quasi-isometric. The asymptotic behavior of the
expectation value of $L$ is therefore similar to the one of
$\left<d(w(\eta))\right>$:
\be
\left\langle L_t^{D}\right\rangle \sim t \quad \mbox{for $t\to\infty$} \label{bound}
\ee
For the bounded domain, the topological complexity grows much faster, as the
infinite space effect discussed above is absent. The scaling
$\left<L_t^{D}\right>\propto t$ should be compared to the (commutative) scaling
$\left<L_{\rm comm}\right>\propto \sqrt{t}$ for two particles. It is noteworthy to
stress that (\ref{bound}) is fully consistent with a discrete model considered in
\cite{voitnech3,raf}.

\section{Conclusion}

To conclude, we would like to comment the new physical content of our results. So
far the only topological invariant studied in the context of entanglements of
fluctuating linear objects was the so-called winding number
\cite{belisle,alain1,nelson,kardar}, which is known to be incomplete for more than
two linear objects. We here propose a model, which describes exactly the underlying
non-Abelian topology of the problem.

We propose a simple geometrical construction of topological invariants of 3--strand
Brownian braids viewed as world lines of 3 particles performing independent Brownian
motions in the complex plane $z$. Our construction is based on the properties of
conformal maps of doubly-punctured plane $z$ to the universal covering surface. We
pay special attention to the case of indistinguishable particles. Our approach is
mainly "self made" and its geometrical transparency we consider as the basic
advantage. The standard machinery of constructing the nontrivial braid group
representations from the Conformal Field Theory is outlined in the Appendix where we
mainly review the strategy realized in \cite{todorov}. The Appendix is added to our
paper exclusively to establish some links between the approachs based on the
geometry of conformal maps and on CFT.

Our method of conformal maps allow us to investigate the statistical properties of
the topological complexity of a bunch of 3--strand Brownian braids and to compute
the expectation value of the irreducible braid length.

\begin{appendix}
\section{Braid group representation and CFT}

In this section we try to describe the derivation of the braid group representation
from the monodromies of some CFT. We will mainly describe the system under
investigation in physical terms, rather than in more rigorous but less transparent
algebraic topological setting.

So, consider $n$ ($n=3$ shall be mainly treated) indistinguishable particles, living
in the complex plane $\C$. The "static" (or configurational) phase space is the
following set $X_n=Y_n/S_n=\C\times Y_{n-1}/S_n$ where
$Y_n=\{{\bf{z}}=(z_1,...,z_n)\in\C^n;\ i\not=j\Rightarrow z_i\not=z_j\}$ and $S_n$
is the group of permutations of $n$ elements. The above defined decomposition means
that one can factor out the coordinate of the center of mass of the system. Thus, we
study the topology of $X_n$, or more precisely of its first homotopy group
$\pi_1(X_n)$, describing the classes of equivalence of closed curves on $X_n$ where
$B_n=\pi_1(X_n)$.

In physical terms we can rephrase the said above as follows. A quantum mechanical
description of this system requires to define the wave functions, or in other words
an $n$ fold tensor--valued functions $ \Psi:X_n\rightarrow V^{\otimes n}$ over this
configuration space. These functions should represent the internal quantum numbers
such as a spin for each particles. The most general case consists in taking $\Psi$
as the so-called $G$--modules, where $G$ is a Lie group. The obtained multi--valued
(in fact, tensor--valued) structure is called a fiber bundle. The main difficulty in
such construction is as follows: the functions $\Psi({\bf{z}})$ and
$\Psi({\bf{z'}}\not= {\bf{z}})$ belong to different spaces, say $h_{\bf z}$ and
$h_{\bf z'}$, and therefore can not be directly compared. To overcome this
difficulty one has to define a map $\disp T_{\gamma_{\bf zz'}}:h_{\bf z}\rightarrow
h_{\bf z'}$ that "transports" the wave function from $\bf z$ to $\bf z'$ along the
path $\gamma$.

Now let us define the holonomy operator, $T_{\gamma_{\bf zz'}}$, and a one--form
$\omega=dT_{{\bf zz}}$, called "flat connection" over $X_n$. The important point is
that the flatness of the connection
$$
d\omega+\omega\wedge\omega=0
$$
is a necessary and sufficient condition for the holonomy group $P_{\bf z}$ at $\bf
z$ to give rise to a {\it monodromy representation} of the fundamental group
$\pi_1(X_n)$. Recall that the holonomy group is the group of $T_{\gamma_{\bf zz}}$
of all closed paths $\gamma_{\bf zz}$ with the natural path composition as an
internal law. We are therefore lead by this statement to a study of flat connections
over $X_n$.

What is the physical meaning to be extracted from such consideration? The basic
idea, put forward by the discovery of the non-Abelian Bohm--Aharnov effect
\cite{verlinde}, is that a closed trajectory in the phase space can affect internal
quantum numbers of the system giving rise to topological interaction which depend
only on the homotopy class of this closed path. The flat connections usually
considered in the literature are the following matrix valued 1--forms, or the
so-called Knizhnik-Zamolodchikov (KZ) connections:
\be
\omega^{KZ}=\sum_{1\le<a<b\le n}\Omega_{ab}d\log z_{ab} \label{eq:connection}
\ee
where $\Omega_{ab}\in {\rm End}(V^{\otimes n})$ is the Casimir invariant (not
explicitly written here), and $z_{ab}=z_a-z_b$. The reader is referred to
\cite{kanie} for a more detailed description of this aspect. This connection was
introduced in 2D conformal field theory in the context of the
Wess--Zumino--Novikov--Witten model and is related to the study of chiral current
algebras \cite{witten,kzam}. In this model, a primary field is covariant under two
kinds of transformations: local gauge transformations generated by the current $J$
and conformal reparameterization generated by the stress-energy tensor $T$. These
$T$ and $J$ are linked by the Sugawara formula describing consistency between two
covariances, and leads hence to the KZ equation for an $n$--points correlator
$\Psi$:
\be
h\,d\Psi=\omega\Psi \label{eq:kz}
\ee
$h$ being a complex parameter. This formulation suggests a geometric interpretation:
$\Psi$ is covariantly constant for this connection. The equation (\ref{eq:kz}) being
a first order linear differential equation, can be treated in the frameworks of the
analytic theory of differential equations. Any solution of (\ref{eq:kz}) can be
represented by a linear combination of its $n-1$ fundamental solutions. The formal
(and in practice not very useful) description of the holonomy group:
\be
T_{\gamma_{\bf zz}}=P\exp\int_{\gamma_{\bf zz}}\omega^{KZ}
\ee
then reduces to a matrix representation. It is shown in \cite{kzam} that equation
(\ref{eq:kz}) can be reduced in the case $n=3$ to a system of ordinary differential
equations for the $SU(N)$--invariant amplitude $\disp F\left(\eta=
\frac{z_{23}}{z_{13}}\right)$ defined by
\be
\Psi(z_1,z_2,z_3)= z_{13}^{-\frac{3}{4h}}(\eta(1-\eta))^{-\frac{N+1}{Nh}}F(\eta)
\ee
This parameterization yields
\be
\left(h\frac{d}{d\eta}+\frac{K_{12}}{1-\eta}-\frac{K_{23}}{\eta}\right) F(\eta)=0
\ee
where $K_{12},\ K_{23}\in{\rm End}(V^{\otimes n})$. Denoting then by $I_0,I_1$ the
basis of $SU(N)$ invariant tensors in $V^{\otimes 3}$ such that
\be
K_{23}I_0=0,\ K_{12}I_1=0
\ee
one can write
\be
F(\eta)=(1-\eta)f^{0}(\eta)I_0+\eta f^{1}(\eta)I_1
\ee
and then reduce (\ref{eq:kz}) to the following system of ordinary differential
equations \cite{todorov}:
\be
\left\{\begin{array}{rcl}
\disp h(1-\eta)\frac{df^0}{d\eta} & = & (h-2)f^0+f^1 \medskip \\
\disp h\eta\frac{df^1}{d\eta} & = & (2-h)f^1-f^0
\end{array}
\right. \label{eq:syst}
\ee
which admits the form of ordinary 2nd order Riemann differential equation with
branching points at $\eta=0,1$
\be
\eta(\eta-1)\frac{d^2f^{j}}{d\eta^2}+\left(\left(3-\frac{4}{h}\right)-1-j+\frac{2}{h}
\eta\right)\frac{df^{j}}{d\eta}+\left(1-\frac{1}{h}\right)
\left(1-\frac{3}{h}\right)f^j=0 \qquad (j=0,1) \label{eq:riem}
\ee
The basis of fundamental solutions of (\ref{eq:riem}) can be written in terms of
standard hypergeometric functions
\be
\begin{array}{ccc}
\disp f^{j}_{0}(\eta)&=&\disp\frac{B(h^{-1},j+h^{-1})}{B(h^{-1},2h^{-1})}
(1-\eta)^{2h^{-1}+j-1}F(j-h^{-1},j+h^{-1},j+2h^{-1},1-\eta) \medskip \\
\disp f^{j}_{1}(\eta)&=&\disp\frac{B(h^{-1},1-j+h^{-1})}{B(h^{-1},2h^{-1})}
\eta^{2h^{-1}-j}F(1-j-h^{-1},1-j+h^{-1},1-j+2h^{-1},\eta)
\end{array} \label{eq:hyper}
\ee
Using (\ref{eq:hyper}) and the integral representations of the hypergeometric
functions, the authors of \cite{todorov} have directly compute the action of $B_3$
generators $\sigma_1,\ \sigma_2$ in the basis of solutions. Choosing a path in the
homotopy class of $\sigma_i$ ($i=1,2$):
\be
\sigma_i:\ z_{\sigma_i}(t)=\frac{z_i+z_{i+1}}{2}+\frac{1}{2}z_{i,i+1}e^{-i\pi t},\
z_{j\not=i,i+1}={\rm const}
\ee
corresponding to an elementary move $\sigma_i$, they end up with the following braid
relations:
\be
\sigma_i:\ f^{j}_{k}\rightarrow [B_i]^{l}_{k}f^{j}_{l}
\ee
where the monodromy matrices are:
\be
B_1={\bar q}^{1/N}\left(\begin{array}{cc} q & 1 \\
0 & -{\bar q} \end{array}\right);\
B_2=q^{-1/N}\left(\begin{array}{cc}-{\bar q} & 0 \\
1 & q \end{array}\right)
\ee
with $q=e^{-\frac{i\pi}{h}}$. This is a 2--dimensional  representation of $B_3$.

Following the general method developed at the length of the Sections \ref{sect:r-h}
and \ref{sect:3}, we can conjecture that
$$
w(\eta)=\frac{f^{j}_{0}(\eta)}{f^{j}_{0}(\eta)}
$$
is a topological invariant of the group $B_3$. In particular, we  expect that $\disp
\frac{dw(\eta)}{d\eta}$ gives the corresponding explicit expression of the
non-Abelian flat connection.

\end{appendix}

\end{document}